\documentclass[12pt,preprint]{aastex}
\newcommand{\nraoblurb}{The National Radio Astronomy Observatory is
a facility of the National Science Foundation operated under cooperative
agreement by Associated Universities, Inc.}
\newcommand{\m}{$\,{\rm m}$}

\newcommand{\khz}{$\,{\rm kHz}$}
\newcommand{\mhz}{$\,{\rm MHz}$}
\newcommand{\ghz}{$\,{\rm GHz}$}
   
\newcommand{\hr}{$\,{\rm hr}$}
\newcommand{\second}{$\,{\rm s}$}
\newcommand{\kms}{${\,{\rm km\, sec^{-1}}}$}

\newcommand{\degree}{\,^\circ}

\newcommand{\arcs}{$^{\prime\prime}$}
\newcommand{\arcmper}{\rlap.{^{\prime}}}

\newcommand{\jy}{\,Jy}
\newcommand{\kel}{\,K}
\newcommand{\mk}{\,mK}
\newcommand{\he}[1]{$^{#1}{\rm He}$}
\newcommand{\hi}{H~{$\scriptstyle {\rm I}$}}
\newcommand{\hii}{H~{$\scriptstyle {\rm II}$}}
\newcommand{\hei}{He~{$\scriptstyle {\rm I}$}}

\newcommand{\oiii}{O~{$\scriptstyle {\rm III}$}}

\newcommand{\hepr}[1]{$^{#1}{\rm He}^{+}/{\rm H}^{+}$}
\newcommand{\her}[1]{$^{#1}{\rm He}/{\rm H}$}

\newcommand{\heri}[1]{${\rm He}^{+}/{\rm H}^{+}$}
\newcommand{\oratio}{${\rm O}^{++}/{\rm O}$}

\newcommand{\ngc}[1]{NGC\thinspace #1}
\newcommand{\sh}[1]{S\thinspace #1}
\newcommand{\Yp}{$Y_{\rm p}$}
\newcommand{\dYdZ}{$\Delta \,Y/\Delta \,Z$}
\newcommand{\dYdO}{$\Delta \,Y/\Delta \,O$}
\catcode`\@=11
\newcommand{\gsim}{${\mathrel{\mathpalette\@versim>}}$}
\newcommand{\lsim}{${\mathrel{\mathpalette\@versim<}}$}
\newcommand{\@versim}[2]{\lower 2.9truept \vbox{\baselineskip 0pt \lineskip
    0.5truept \ialign{$\m@th#1\hfil##\hfil$\crcr#2\crcr\sim\crcr}}}
\catcode`\@=12

\begin{document}

\title{The Chemical Evolution of Helium}

\shortauthors{Balser 2005}

\author{Dana S. Balser\altaffilmark{1}}

\altaffiltext{1}{National Radio Astronomy Observatory,
P.O. Box 2, Green Bank WV 24944, USA.}

\begin{abstract}

We report on measurements of the \he4\ abundance toward the outer
Galaxy \hii\ region \sh{206} with the NRAO Green Bank telescope.
Observations of hydrogen and helium radio recombination lines between
$8-10$\ghz\ were made toward the peak radio continuum position in
\sh{206}.  We derive \her4\ = $0.08459 \pm\ 0.00088\, {\rm (random)}
\pm\ 0.0010\, {\rm (known \,\, systematic)}$, 20\% lower than optical
recombination line results.  It is difficult to reconcile the large
discrepancy between the optical and radio values even when accounting
for temperature, density, and ionization structure or for optical
extinction by dust.  Using only M17 and \sh{206} we determine \dYdZ\ =
$1.41 \pm\ 0.62$ in the Galaxy, consistent with standard chemical
evolution models.  High helium abundances in the old stellar
population of elliptical galaxies can help explain the increase in UV
emission with shorter wavelength between 2000 and 1200\,\AA, called
the UV-upturn or UVX.  Our lower values of \dYdZ\ are consistent with
a normal helium abundance at higher metallicity and suggest that other
factors, such as a variable red giant branch mass-loss with
metallicity, may be important.  When combined with \he4\ abundances in
metal poor galaxy \hii\ regions, Magellanic cloud \hii\ regions, and
M17 that have been determined from optical recombination lines
including the effects of temperature fluctuations, our radio
\her4\ abundance ratio for \sh{206} is consistent with a helium
evolution of \dYdZ\ = 1.6.  A linear extrapolation to zero metallicity
predicts a \her4\ primordial abundance ratio about 5\% lower than that
given by the {\it Wilkinson Microwave Anisotropy Probe} and standard
Big Bang nucleosynthesis.  The measured \he4\ abundances may be
systematically underestimated by a few percent if clumping exists in
these \hii\ regions.

\end{abstract}


\section{Introduction}

Helium is the second most abundant element in the Universe.  Most of
the \he4\ is thought to have been produced during the era of primordial
nucleosynthesis several minutes after the Big Bang \citep{peebles66}.
Measurements of the primordial \her4\ abundance ratio have been made
by observing optical emission lines in low metallicity extragalactic
\hii\ regions and either taking the average value of \her4\ in the 
most metal poor objects (e.g., Searle \& Sargent 1972) or
extrapolating back to zero metallicity (e.g., Peimbert \&
Torres-Peimbert 1974).  Dwarf irregular and blue compact galaxies have
been extensively studied \citep[and references therein]{it04}.  More
recently, the primordial helium abundance has been determined by using
cosmic microwave background (CMB) observations from the {\it Wilkinson
Microwave Anisotropy Probe (WMAP)} to measure the baryon-to-photon
ratio along with standard Big Bang nucleosynthesis (SBBN) calculations
\citep{romano03, cyburt03, coc04}.  Depending on the adopted
uncertainties the measured \he4\ abundance is either in agreement with
WMAP/SBBN or lower than expected \citep{it04, olive04, steigman05}.

The only other significant known source of helium is the synthesis of
\he4\ in stars \citep{burbidge57}.  The rate of \he4\ production is
typically expressed relative to the change in metallicity, \dYdZ,
where $Y$ and $Z$ are the helium and metal abundances by mass.
The functional form of \dYdZ\ is determined by observations of \he4\ in
objects with different metallicities.  A variety of sources have been
used: extragalactic \hii\ regions \citep{peimbert74, pagel92,
skillman94, peimbert00, it04}, Galactic \hii\ regions \citep{shaver83,
peimbert88, baldwin91, deharveng00}, Galactic planetary nebula
\citep{dodorico76, maciel01}, K dwarf stars \citep{pagel98,
jimenez03}, and the Sun \citep{basu95, share97, bahcall05a, bahcall05b}.

The primary diagnostic for direct measurements of \he4\ has been optical
recombination lines (ORLs).  There are multiple, bright transitions of
H and \he4\ in \hii\ regions and planetary nebulae (PNe).  Moreover,
transitions of heavier elements can be used to probe the ionization
structure and metallicity.  The observed H and \he4\ line intensities
must be converted into a \her4\ abundance ratio.  Sources of
uncertainty in these calculations include (1) underlying absorption
lines from stellar emission; (2) collisional excitation of \hi\ and
\hei\ lines; (3) \hei\ optical depth effects; and (4) density, temperature
and ionization structure \citep{olive01, luridiana03, peimbert03b}.

Radio recombination lines (RRLs) are weaker than ORLs but their
interpretation should be simpler.  The high principal quantum number
(n) states of hydrogen and helium should be altered by radiative and
collisional effects in the same way so that the ratio of the
intensities is equal to the abundance ratio.  There still exists the
ionization structure problem since there is no way to directly measure
neutral helium.  Early RRL observations of Galactic \hii\ regions,
however, produced \hepr4\ abundances that varied with n
\citep{lockman82}.  (See Roelfsema \& Goss [1992] for a
review of high spatial resolution RRL observations with
interferometers.)  This effect was attributed to (1) departures from
LTE that would effect H and \he4\ differently \citep{baldwin91}; (2) a
geometric effect that assumes the degree of helium ionization
decreases from the center of the \hii\ region and that observations
with higher n have lower spatial resolution \citep{mezger80}; and (3)
that poor spectral baselines would systematically effect either the
pressure broadened high-n transitions or the weaker helium lines due
to confusion with weak \hii\ emission along the line of sight
\citep{lockman82}.  \citet{peimbert92a} used $\alpha$, $\beta$, and 
$\gamma$ RRL transitions at the same spatial resolution and determined \hepr4\
abundance ratios consistent within the errors.  These data had better
sensitivity and improved spectral baselines over previous RRLs
measurements \citep{balser94}.  Nevertheless, the uncertainty in
\hepr4\ was around 10\% or larger.

We report here sensitive H and \he4\ radio recombination line
observations toward the \hii\ region \sh{206} to compare with ORL
results.  This outer Galaxy \hii\ region is ionized by an O4-O5 star
and is the only object in a survey of 36 high excitation \hii\ regions
that contains no neutral helium \citep{deharveng00}.  Optical
recombination lines of H and \he4\ were made with a Fabry-Perot
spectrophotometer and \hepr4\ abundances were calculated for different
positions, diaphragm sizes, and determined electron temperatures.  The
average value of those listed in their Table 1 is \hepr4$ = 0.1036 \pm\
0.0027$.  Recent RRL data for \sh{206} yeilds \hepr4$ = 0.0924
\pm\ 0.0080$ \citep{quireza06}.  The quoted uncertainties are based on
the formal errors of Gaussian fits to the line profiles.  Instrumental
spectral baseline effects will produce uncertainties that can be as
large as the random errors \citep{balser94, quireza06}.  Improved RRL
data would provide an important independent constraint on the ORL
results.

\section{Observations}

The observations were made at X-band ($8-10$\ghz) with the National
Radio Astronomy Observatory\footnote{\nraoblurb} 100\m\ Green Bank
telescope (GBT) in June 2004.  The GBT has an unblocked aperture that
significantly reduces reflections from various parts of the telescope
structure.  These reflections produce standing waves that limit the
detection or accurate measurement of weak, wide spectral lines
\citep{balser94, bania97}.  The Sharpless \hii\ region S206 was
observed at the J2000 position (right ascension = 04:03:15.87,
declination = +51:18:54) assuming an LSR (radio) velocity of
$-25.4$\kms.  The half-power beamwidth (HPBW) of the GBT is 80\arcs\
at a frequency of 9\ghz.

The flux density scale was measured by injecting noise into the
signal path of a known intensity as a function of frequency.  This
intensity scale was verified using the astronomical calibration source
\ngc{7027} which has a flux density of 6.0\jy\ between $8-10$\ghz\
\citep{peng00}.  These two methods of calibration are consistent
to within 5-10\%.  Local pointing and focus corrections were made
approximately every two hours using nearby pointing sources
\citep{condon01}.

Continuum data were taken at a frequency of 9\ghz\ with a bandwidth of
320\mhz\ using the digital continuum receiver.  A Nyquist sampled
continuum image of \sh{206} was produced by driving the telescope back
and forth in right ascension at a rate of 60\arcmin\ per minute for
60\second.  The image consists of 121 raster scans offset by
$0\arcmper5$ in declination.  An integration time of 0.1\second\
provided $0\arcmper1$ sampling in right ascension.  The continuum data
were analyzed using AIPS++.  A linear baseline was fit to the outer
10\% of each raster scan and subtracted from the data.  The two
orthogonal circular polarizations were averaged.  The data were placed
on a uniform grid using a prolate spheroidal convolving function.

Spectra were taken using the total power position switching mode where
a reference position (OFF) was observed offset $\sim 6$ minutes in
right ascension from the source and then the target position (ON) was
observed.  The reference and target positions were observed for 6
minutes each for a total of 12 minutes.  The GBT Spectrometer was
configured with 8 spectral windows times 2 orthogonal circular
polarizations (LCP and RCP) for a total of 16 independent spectra.
Each spectrum consisted of 4096 channels with a bandwidth of 50\mhz\
yielding a frequency resolution of 12.2\khz.  All spectra have been
Hanning smoothed.

Table~\ref{tab:rrl} summarizes the H radio recombination lines that
were observed.  Listed are the Hn$\alpha$ transition, the Hn$\alpha$
RRL frequency, the corresponding velocity resolution, the spectral
window center frequency, and higher order H RRLs within the 50\mhz\
bandwidth.  The corresponding He RRLs are shifted $\sim 3-4$\mhz\
higher in frequency.  The frequency at the center of the band was
typically offset in frequency from the Hn$\alpha$ transition to
observe higher-order RRLs that can be used to model the physical
properties of \sh{206} and monitor system performance (e.g., Balser et
al. 1999).

The spectral line data were calibrated and analyzed using an IDL
software package created for single-dish data reduction.\footnote{See
http://www.bu.edu/iar/research/dapsdr/.} Each spectrum was visually
inspected and a very small fraction of the data were edited due to
interference.  Because RRLs involve high-n transitions the expected
line profiles should be similar between adjacent lines.  Therefore the
Hn$\alpha$, Hen$\alpha$, and Cn$\alpha$ transitions can be averaged to
increase the signal-to-noise ratio.  The velocity resolution of each
spectral window is different, however, because they are centered at
different sky frequencies.  Moreover, the spectra are sampled at
different velocities.  Therefore all spectra were re-sampled onto the
87$\alpha$ band using a sin(x)/x interpolation \citep{roshi05}.  The
86$\alpha$ band was not used because of confusion with higher order
RRLs.  Therefore there were 14 independent spectra, 7 RRLs times 2
polarizations, each with a typical $rms$ noise of $\sim 1$\mk.  The
expected $rms$ noise of the averaged spectrum is $\sim 1/\sqrt{14}$\mk =
0.27\mk, very close to the measured $rms$ noise of 0.28\mk.

\section{Results}

\begin{figure}
\includegraphics[angle=0, scale=0.75]{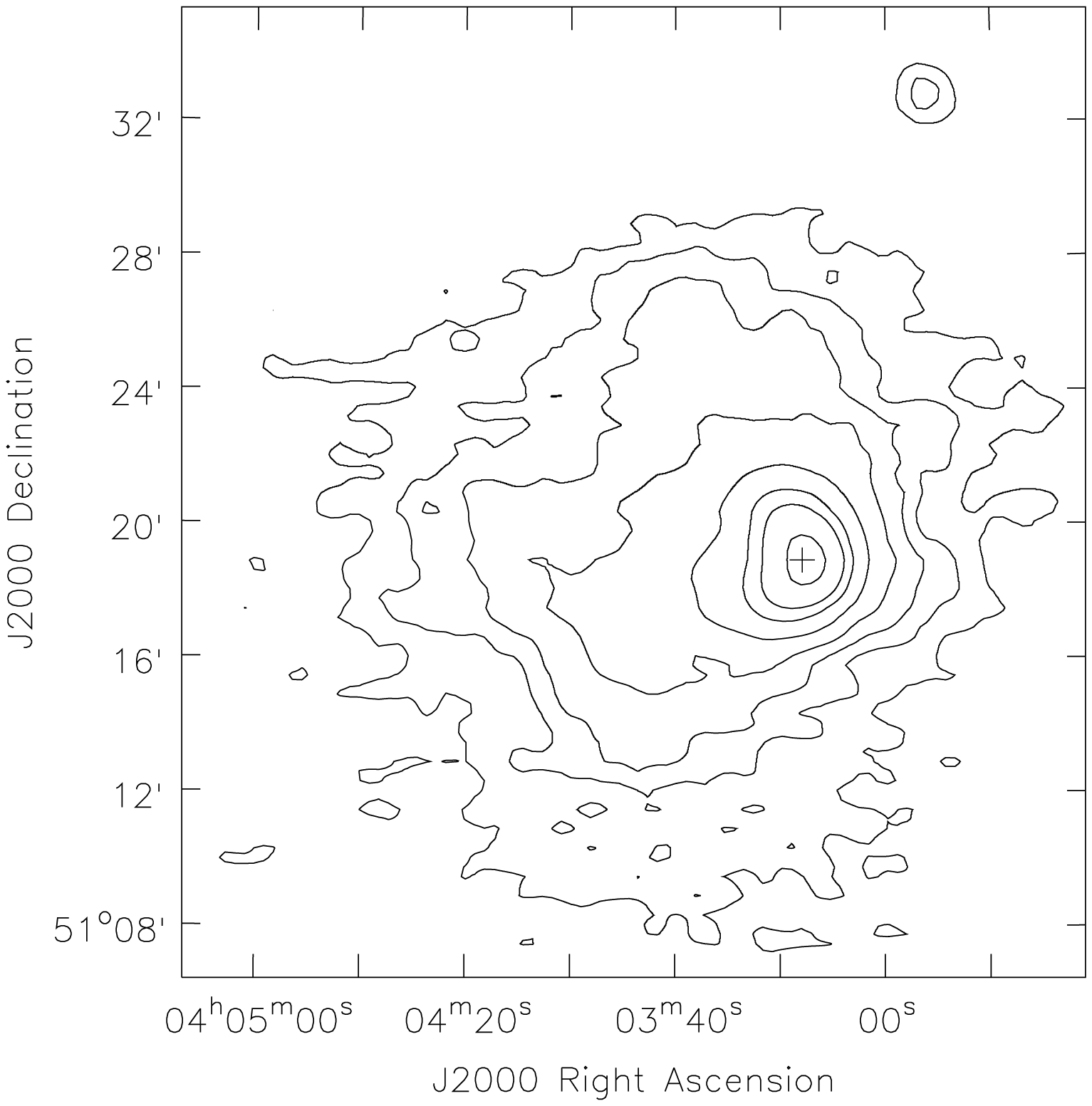}
\figcaption{Continuum image of \sh{206} at a frequency of 9.0\ghz.
The intensity scale is the antenna temperature in units of Kelvin.
The peak intensity in the image is 2.86\kel.  The $rms$ noise of the
image is 14.1\mk.  The contour levels are at 1, 2, 3, 5, 10, 20, 30,
and 50 times the $3 \sigma$ (42.3\mk) level.  The plus symbol
indicates the position of the GBT radio recombination line
observations.
\label{fig:cont}}
\end{figure}

The GBT continuum image of \sh{206} is shown in Figure~\ref{fig:cont}.
This relatively low resolution radio continuum image suggest a
core/halo morphology (c.f., Walmsley et al. 1975; Balser et al. 1995).
Higher resolution radio observations reveal structure within the core
component \citep{deharveng76, albert86, balser95, omar02}.  Optical
H$\alpha$ images of \sh{206} are similar in morphology and contrast
to the radio continuum images, suggesting no high extinction dust
clouds \citep{deharveng76}.  The major source of ionizing radiation is
BD+50$\degree$886 which has been classified as either an O4 or O5V
star \citep{crampton74, hunter90}.  The ionizing star is located about
one arcmin to the east of the brightest emission, consistent with a
blister-type model with the \hii\ region being mass-limited to the
east and photon-limited to the west.  Infrared studies support this
general picture \citep{albert86, pismis91, mookerjea99}.

\begin{figure}
\includegraphics[angle=90, scale=0.5]{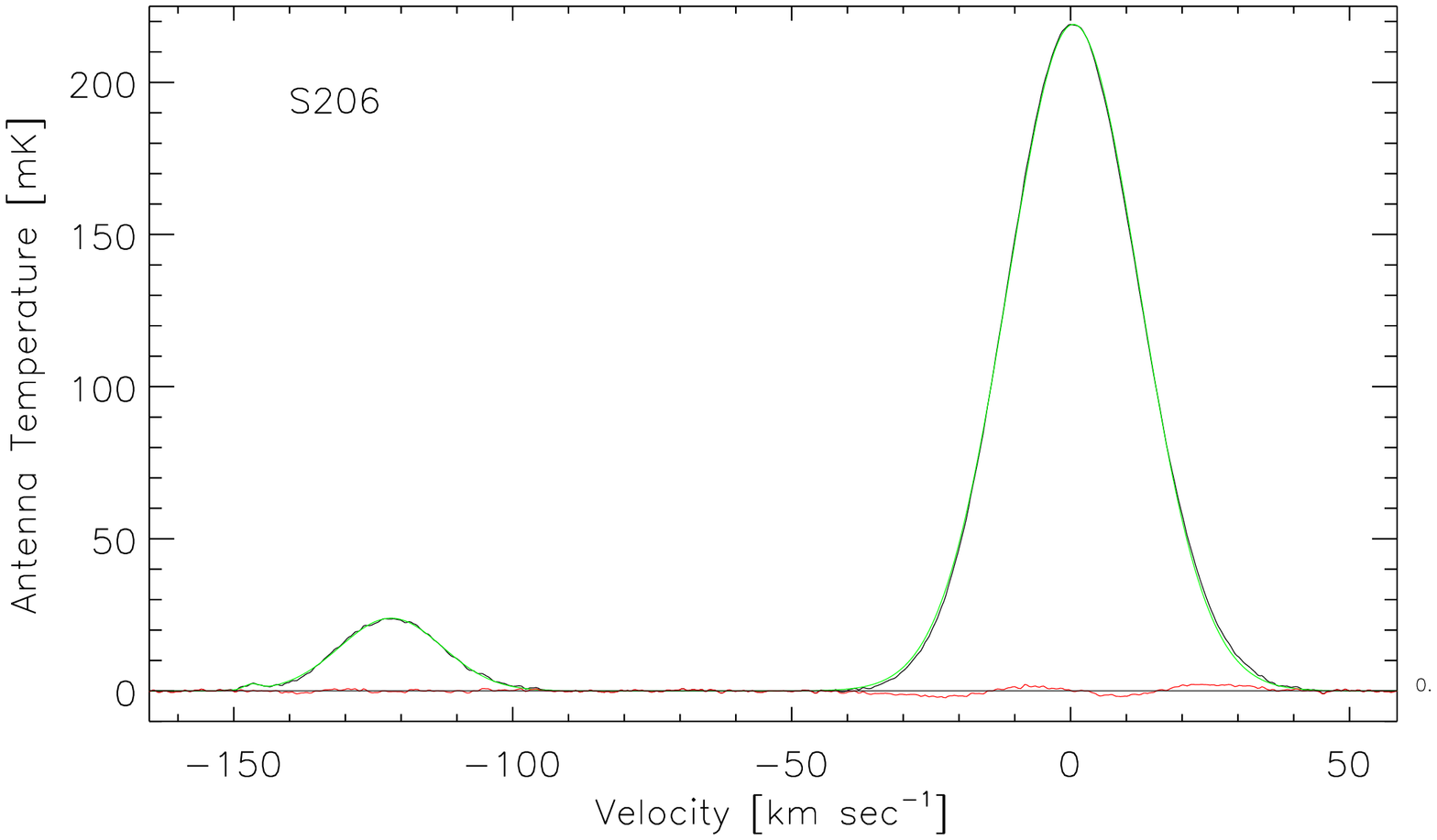}
\includegraphics[angle=90, scale=0.5]{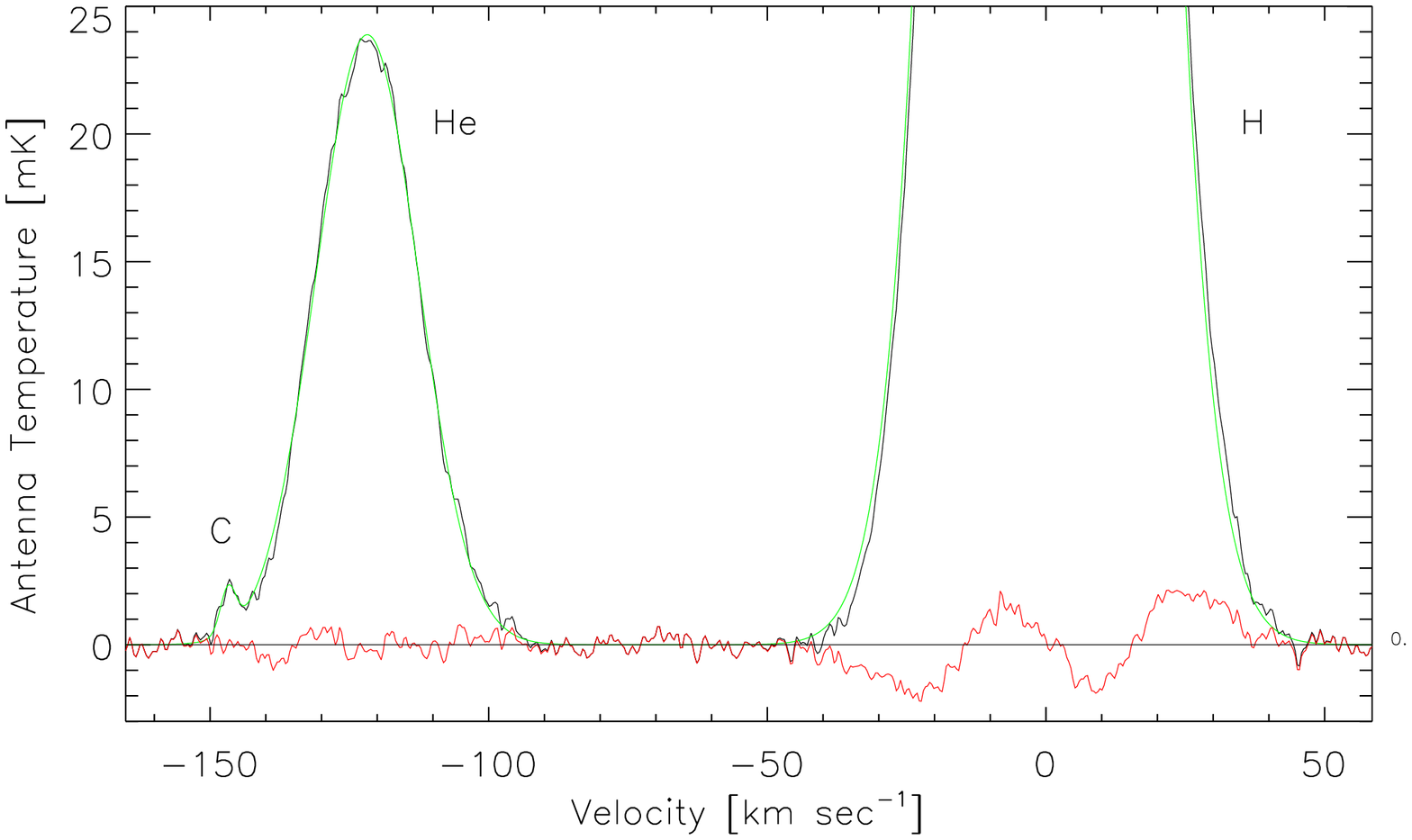}
\figcaption{Averaged spectrum of \sh{206} consisting of the
$87\alpha - 93\alpha$ transitions.  The 86$\alpha$ transition has not
been included (see text).  The intensity scale is the antenna
temperature in units of milliKelvin.  All spectra were interpolated to
the $87\alpha$ velocity scale before averaging.  The black curves are
the data, the green curves are Gaussian models, and the red curves are
the residuals.  The velocity at the Hn$\alpha$ line center has been
arbitrarily set to zero.  The $rms$ noise in the line-free regions is
0.28\mk\ for a total integration time of 555.3\hr.  The bottom panel
shows an expanded scale.
\label{fig:line}}
\end{figure}

The averaged alpha-line spectrum is shown in Figure~\ref{fig:line}.
The H and \he4\ transitions are detected with high signal-to-noise
ratios.  Shown are Gaussian fits to the data along with the residuals.
The data are well modeled by a single Gaussian function; although
there is some evidence of non-Gaussian structure in the residuals for
the H and He transitions.  Table~\ref{tab:results} summarizes the
results of the Gaussian models.  Listed are the peak intensity and the
full-width half-maximum linewidths, along with their associated
errors.

Although the GBT spectral baselines are much improved over
traditionally designed single-dish telescopes, there does exist some
baseline structure.  Therefore a low-order ($2^{\rm nd}$ or $3^{\rm
rd}$) polynomial model was fit to the line-free baseline regions of
each alpha-line spectrum and subtracted from the data.  This
removes the continuum level and any instrumental baseline structure
before the data are re-sampled to the 87$\alpha$ velocity scale.  The
measured $rms$ noise of the line-free regions is consistent with the
theoretically predicted value.  The calculated \hepr4\ abundance
ratio, based on Gaussian models, is $0.08459 \pm\ 0.00088$.  The
uncertainty is based on the formal errors of the Gaussian model.  The
\hepr4\ abundance ratio is proportional to the ratio of the He and H
RRL intensity.  Therefore, the uncertainty of $5-10$\% in the absolute
intensity scale cancels.  We tested this by averaging the LCP and RCP
spectra separately.  The \hepr4\ ratios for LCP, RCP, and the combined
(LCP plus RCP) are consistent to within the random uncertainties.  We
estimate a systematic error of about 0.1 percent based on different
baseline models.  The helium within the \hii\ region of \sh{206}
appears to be fully ionized; therefore we determine \her4\ = $0.08459
\pm\ 0.00088\, {\rm (random)} \pm\ 0.0010\, {\rm (known \,\, systematic)}$.

\section{Discussion}

The \her4\ abundance ratio determined using RRLs in \sh{206} is 20\%
lower than values determined from ORLs.  Direct comparison of
\her4\ using radio and optical techniques is often difficult because the two
methods usually do not probe the same region.  Therefore any true
variation in \her4\ or ionization structure will complicate the
comparison.  For example, ORLs are typically observed using a long
slit across the nebula while RRLs are observed over a larger area,
especially when single-dish telescopes are used.  Also, dust can cause
significant extinction at optical wavelengths whereby only the near
side of the nebula is probed, while RRLs at cm-wavelengths are
sensitive to the entire volume within the telescope's resolution or
beam.  Nevertheless, the ORLs discussed in \citet{deharveng00} were
observed with a Fabry-Perot for different positions and diaphragm
sizes that covers the GBT's beam.  Also, infrared and optical images
of \sh{206} indicate that there are no high extinction dust clouds
\citep{deharveng76, albert86, pismis91, mookerjea99}.  Therefore, the
radio and optical data should sample a similar volume of the nebula.
The \hepr4\ ratio is constant to within the uncertainties throughout
the nebula while the \oratio\ ratio varies from $\sim 0.70-0.85$
\citep{deharveng00}.  Therefore \sh{206} should contain no neutral helium.

\hepr4\ abundance ratios calculated using ORLs are a weak function of the 
electron temperature \citep{peimbert95}.  Electron temperatures
determined from collisionally excited lines (CELs) are sensitive to
higher temperature regions while recombination line and recombination
continua are sensitive to lower temperature regions.  Therefore,
temperature inhomogeneities can be probed using at least two of these
different temperature diagnostics. \citet{peimbert67} developed a
formalism to determine the effects of temperature fluctuations using
two parameters: the average temperature ($T_{\rm o}$) and the mean
squared temperature fluctuation ($t^2$).  For example, \hepr4\
abundances determined using ratios of CELs of [\oiii] can overestimate
the electron temperature and thus the helium abundance
\citep{peimbert69, peimbert74, peimbert95, peimbert02b}.  Even for
uniform electron temperature nebulae, density fluctuations can alter
the emissivities of CELs that are used to determine the electron
temperature \citep{rubin89, viegas94}.

For \sh{206}, \citet{deharveng00} used electron temperatures from the
[\oiii] CEL ratios where $T_{\rm e} = 9016 \pm\ 203$\kel\ is the mean
of their individual temperature measurements.  Models of \sh{206}
using RRL and continuum emission including non-LTE effects determine
an electron temperature of $9000 \pm\ 500$\kel\ \citep{balser99}.  For
the GBT data here we calculate an LTE electron temperature of $T_{\rm
e} = 8856 \pm\ 39$\kel\ using the line and continuum data
\citep{mezger67}.  Therefore it does not appear that temperature or
density fluctuations can account for the large differences in \hepr4\
determined from ORLs and RRLs.

\citet{liu00} have proposed that cold, H-deficient clumps exist in
nebulae that dominate the ORL emission.  Since the H and He
recombination lines are more heavily weighted toward the cold, dense
clumps, the measured \hepr4\ abundance will be overestimated compared
to the average \hepr4\ value.  But ORLs and RRLs have the same
dependence on density and therefore should be affected in the same
way.  Thus, it seems difficult to explain the helium abundance
discrepancy between the optical and radio results for \sh{206}.

\begin{figure}
\includegraphics[angle=270, scale=0.55]{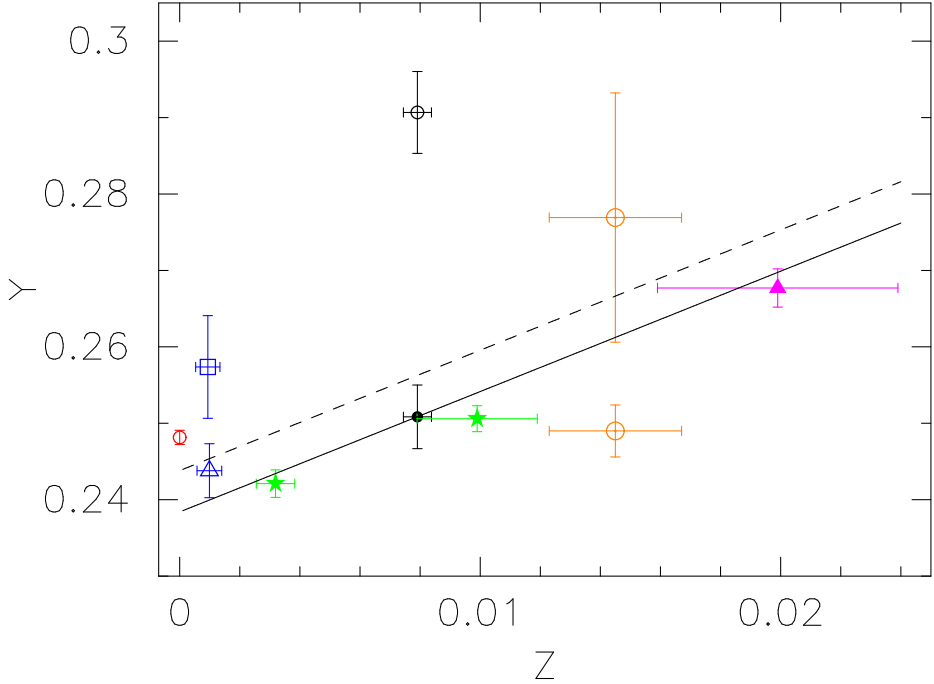}
\figcaption{Plotted is the \her4\ abundance ratio by mass ($Y$)
versus the metallicity ($Z$).  The red open circle is
the primordial helium abundance (\Yp) determined from WMAP observations
and standard Big Bang nucleosynthesis \citep{spergel06}.  The blue
open symbols are based on a study of 7 blue compact galaxies
\citep{it04}.  The open triangle denotes the mean helium abundances
from \citet{it04} while the open square is the mean helium abundances
from a reanalysis of the same data from \citet{olive04}.  The error
bars are the standard deviations from the mean values.  The green
filled stars are the helium abundances in the lower metallicity SMC
\hii\ region \ngc{346} and the higher metallicity LMC \hii\ region 30
Doradus \citep[and references therein]{peimbert03a}.  The orange
symbols are helium abundances in the Sun using theoretical stellar
evolution models \citep{grevesse96} and from helioseismology
\citep{basu04}.  The Solar metallicity is taken from 
\citet{peimbert03a}.  The helium abundances determined from 
helioseismology predicts lower values of $Y$.  The filled magenta
triangle at higher metallicity is the helium abundance for the
Galactic \hii\ region M17 \citep[and references therein]{peimbert03a}.
The black circles are the helium abundances for the Galactic \hii\
region \sh{206} using ORLs from \citet{deharveng00} denoted by the
open circle and RRLs in this paper (filled circle).  The solid line is
based on a determination of the primordial helium abundance using
metal poor galaxies with \Yp\ = $0.2384 \pm\ 0.0025$ and assuming
\dYdO\ = 3.5 or \dYdZ\ = 1.6 \citep{peimbert02b}.  The dashed line
corresponds to a systematic increase of 3\% in the \her4\ abundance.
\label{fig:y_vs_o}}
\end{figure}

Figure~\ref{fig:y_vs_o} plots the \her4\ abundance ratio by mass ($Y$)
versus the metallicity ($Z$) for both Galactic and extragalactic
sources.  For \sh{206} we convert the \her4\ abundance ratio by number
($y$) to $Y$ using $Y = 4y(1 - Z)/(1 + 4y)$ (e.g., Pagel et al. 1992).
We use the oxygen abundance as a proxy for $Z$ and assume that 45\% of
metals is comprised of oxygen by mass (i.e., $O = 0.45Z$; Maciel
[2001]).  The oxygen abundance for \sh{206} is taken from
\citet{deharveng00} where we have increased the abundance by 0.08 dex
to correct for depletion in dust.  We convert the O/H abundance by
number ($o$) to the oxygen abundance by mass ($O$) using $O/(1 - Z +
O) = 16o/(1 + 4y + 16o)$.

Accurate \her4\ abundance ratios are difficult to measure in the
Galaxy.  Helium cannot be directly measured in the Sun and is
typically inferred from other observables.  In Figure~\ref{fig:y_vs_o}
we plot the Solar helium abundance determined from stellar evolution
models \citep{grevesse96} and from helioseismology \citep{basu04}.
New atmospheric models of the Sun predict lower mass fractions for the
heavy elements that are inconsistent with helioseismology and produce
lower values of $Y$ \citep{bahcall05a}.  Other measurements of \her4\
in the Galaxy include ORLs and RRLs in \hii\ regions and planetary
nebulae (PNe).  The major problem with such studies is to determine
the total \her4\ abundance ratio.  This requires a correction for any
neutral helium that resides within the \hii\ region since neutral
helium cannot be directly observed.  The Orion nebula is the best
studied, nearby Galactic \hii\ region.  Two distinct values of \her4\
have been determined in Orion based on both optical and radio
observations: $\sim 0.088 \pm\ 0.007$ \citep{baldwin91, pogge92} and
$\sim 0.10 \pm\ 0.008$ \citep{peimbert88, rubin91, esteban99a}.  The
different abundances arise primarily from differences in the
ionization correction for neutral helium.  The lower abundance is
based on an analysis that predicts very little neutral helium.
Because of these discrepancies, \citet{peimbert93} suggested that M17
is a better source to use for measuring \her4\ because the radiation
field should produce little neutral helium.  We plot the helium
abundance of $Y = 0.2677 \pm\ 0.0025$ for M17 \citep{peimbert02a,
peimbert03a}.  This \her4\ abundance ratio is based on a reanalysis of
the data from \citet{esteban99b} using a self consistent method that
includes temperature fluctuations \citep{peimbert02a}.  The new value
of $Y$ is slightly smaller.  Radio measurements of \hepr4\ in M17 are
typically in good agreement with the optical results
\citep{peimbert88, peimbert92a, peimbert92b, peimbert93, esteban99b,
tsivilev99}.  Nevertheless, the radio observations are usually
dominated by systematic effects that produce nonrandom frequency
structure in the instrumental baselines \citep{balser94, bania97}.
Also, most of the RRL data have been observed toward the brighter
south-west component (M17S), while the optical data are from the
north-east (M17N). \citet{quireza06} have observed RRLs toward M17N,
very close to the M17-3 position of \citet{peimbert92a}, and determine
$Y = 0.2688 \pm\ 0.0088$, consistent with the optical result.

Using only M17 and \sh{206} to measure the chemical evolution of
helium we calculate \dYdZ\ = $1.41 \pm\ 0.62$ in the Galaxy.  Chemical
evolution models are consistent with \dYdZ\ $\sim 1$ although higher
values can be produced by enriched supernovae winds (see below).
Other recent estimates of helium production from Galactic \hii\
regions determine \dYdZ\ $\geq\ 2$ \citep{esteban99b, deharveng00,
maciel01}.  Local estimates of \dYdZ\ in the Galaxy have been made by
using the fine structure in the main sequence of nearby stars with
Hipparcos data.  The estimates range from \dYdZ\ $= 2-3$
\citep{pagel98, jimenez03}.

Measurements of \dYdZ\ are also important in understanding the excess
of UV emission observed in elliptical galaxies which may be a useful
diagnostic of age \citep{yi99}.  Observations of UV radiation toward
elliptical galaxies measure an increase in UV emission with shorter
wavelengths between 2000 and 1200\,\AA\ that has been called the
UV-upturn or UVX (see O'Connell [1999] for a review).  Since it's
discovery over 25 years ago the interpretation of the UVX has been
controversial.  Recent evidence suggests that the source of the UVX is
the old stellar population of early-type galaxies \citep{greggio90,
dorman93, dorman95, yi97, yi98}.  In particular, extreme horizontal
branch stars and their progeny are thought to be the main candidates
for the UVX.  Higher helium abundances will produce larger UV emission,
although the adopted mass-loss on the red giant branch is very
important \citep{dorman95, yi98}.  \citet{yi98} prefer a moderate
increase in helium (\dYdZ\ \gsim\ 2) to explain the UVX, while
\citet{dorman95} prefer a smaller helium abundance.  Our observations
of \dYdZ\ in the Galaxy suggest smaller helium production at higher
metallicities.

\hii\ regions in the Magellanic clouds provide a measure of \her4\ at
intermediate metallicities.  In Figure~\ref{fig:y_vs_o}, we plot $Y$
determined from ORLs including corrections for temperature
fluctuations for \ngc{346} in the SMC \citep[and references
therein]{peimbert00, peimbert03a} and 30 Doradus in the LMC
\citep{peimbert03a}.  Observations of \her4\ in the Magellanic clouds
have the advantage that the \hii\ regions are relatively metal poor
and the nebulae can be spatially resolved from the ionizing stars.

There has been a significant effort to measure \he4\ in very low
metallicity objects, such as blue compact galaxies, where a more
direct estimate of the primordial helium abundance (\Yp) can be made.
Although it is difficult to resolve individual \hii\ regions for these
more distant objects the lower metallicity produces a very hard
radiation field that should ionize all of the helium.  In
Figure~\ref{fig:y_vs_o} we show the mean $Y$ values determined from 7
metal poor dwarf galaxies from the original analysis \citep[\Yp\ = $0.2421
\pm\ 0.0021$]{it04} and a reanalysis of the same data \citep[\Yp\ =
$0.249 \pm\ 0.009$]{olive04}.  Using 5 metal poor galaxies
\citet{peimbert02b} determine $Y$ values that are significantly lower
than previous determinations when including the effects of temperature
fluctuations.  Assuming \dYdO\ = 3.5 they determine \Yp\ = $0.2384
\pm\ 0.0025$.  These results are summarized as a solid line in
Figure~\ref{fig:y_vs_o} where we have assumed that oxygen contains
45\% of the metals by mass or \dYdZ\ = 1.6.  The primordial helium
abundance from WMAP/SBBN is also shown with \Yp\ = $0.24815 \pm\
0.00033\, {\rm (random)} \pm\ 0.0006\, {\rm (systematic)}$
\citep{spergel06}.

The \her4\ abundances for \sh{206} are shown in
Figure~\ref{fig:y_vs_o} for both the optical and radio data.  Our
radio $Y$ value is consistent with the optically determined values of
$Y$ for metal poor galaxy \hii\ regions, Magellanic \hii\ regions, and
M17 when temperature fluctuations are included and a constant helium
evolution of \dYdZ\ = 1.6 from stars.  A value of \dYdO\ = $3.5 \pm\ 0.9$ was
adopted by \citet{peimbert00} based on results from different chemical
evolution models of irregular galaxies and the Galaxy \citep{carigi95,
carigi99, chiappini97, carigi00}.  Since the range of metallicities in
metal poor galaxies is small it is difficult to empirically determine
\dYdZ\ from these observations.  Chemical evolution models are in
general consistent with a linear relationship between $Y$ and $Z$,
although they are sensitive to stellar yields \citep{pilyugin93,
fields98}.  Generic models produce \dYdZ\ $\sim\ 1$ while \he4\
enrichment from supernovae winds can increase this value
\citep{marconi94, fields96, copi97, fields98, chiappini02}.  

\cite{olive04} concluded that the uncertainties in determining \her4\ from
ORLs in extragalactic objects have been underestimated.  They suggest
a more conservative range for the primordial helium abundance of
$0.232 < Y_{\rm p} < 0.258$ based on the analysis of ORLs from metal
poor galaxies.  This range of values for \Yp\ covers the results from
other studies, including the WMAP/SBBN prediction.  Nevertheless, it
is interesting that all of the filled symbols in
Figure~\ref{fig:y_vs_o} which include results from ORLs that have been
corrected for temperature fluctuations and our \sh{206} RRL result are
consistent with a single value for \dYdZ\ and chemical evolution
models.  These \her4\ abundances predict a primordial value that is
lower than that given by WMAP/SBBN \citep{peimbert00, peimbert03a}.

\citet{mathis05} have produced Monte Carlo photo-ionization models
of \hii\ regions that include a hierarchical density distribution of
clumps.  They predict that even for very hot stars ($T = 45,000$\kel)
there exists some neutral helium within the \hii\ region for models
that include clumping.  A sharp increase in density causes hydrogen to
compete more effectively for He-ionizing photons; hence He$^{+}$/He
decreases.  They estimate an increase of $\sim 3$\% in
\her4\ to correct for clumping.  This would increase $Y$ for both ORLs
and RRLs studies depending on the amount of clumping in each \hii\
region.  The dashed line in Figure~\ref{fig:y_vs_o} corresponds to a
3\% increase in \her4\ relative to the solid line.  A 5\% increase
will reconcile the lower \her4\ abundances in Figure~\ref{fig:y_vs_o}
defined by the solid line with WMAP/SBBN.

\section{Conclusion}

The following are the main results of our observations with the NRAO
Green Bank telescope of hydrogen and helium radio recombination lines
between $8-10$\ghz\ toward the outer Galaxy \hii\ region \sh{206}:

\begin{enumerate}

\item We determine a \her4\ abundance ratio of $0.08459 \pm\ 0.00088\,
{\rm (random)} \pm\ 0.0010\, {\rm (known \,\, systematic)}$ in
\sh{206}.  The GBT data are of high quality.  The unblocked aperture
of the GBT has significantly reduced the systematic effects of
traditionally designed single-dish telescopes.  Moreover, the GBT
Spectrometer can tune to 8 RRLs simultaneously.  These independent
spectra were averaged to increase the signal-to-noise ratio.

\item Our RRL measurement of \her4\ in \sh{206} is 20\% lower
than ORL results \citep{deharveng00}.  Although temperature and
density fluctuations can result in an overestimate of optically
determined values of \hepr4, these effects appear to be small.  ORLs
of hydrogen and helium along with CELs of oxygen indicate no neutral
helium in \sh{206}.  There is also no evidence of significant dust
extinction suggesting that the optical and radio emission lines are
probing the same volume of gas.  Therefore, it is difficult to explain
the helium abundance discrepancy between the optical and radio results
for \sh{206}.

\item Using only M17 and \sh{206} we determine \dYdZ\ = $1.41 \pm\
0.62$ in the Galaxy, consistent with standard chemical evolution
models.  High helium abundances in the old stellar population of
elliptical galaxies can help explain the increase in UV emission with
shorter wavelength between 2000 and 1200\,\AA, called the UV-upturn or
UVX.  Our lower values of \dYdZ\ are consistent with a normal helium
abundance at higher metallicity and suggest that other factors, such
as a variable red giant branch mass-loss with metallicity, may be
important.

\item Our \her4\ abundance for \sh{206} is consistent with
a helium evolution of \dYdZ\ = 1.6 when combined with metal poor
galaxy \hii\ regions, Magellanic cloud \hii\ regions, and M17 that
have been determined with ORLs including the effects of temperature
fluctuations.  These data are consistent with a primordial helium
abundance of \Yp\ = $0.2384 \pm\ 0.0025$ \citep{peimbert02b},
significantly lower than the WMAP/SBBN value of \Yp\ = $0.24815
\pm\ 0.00033\, {\rm (random)} \pm\ 0.0006\, {\rm (systematic)}$
\citep{spergel06}.  The measured \he4\ abundances may
be systematically underestimated by a few percent if clumping exists
in these \hii\ regions \citep{mathis05}.

\end{enumerate}

\acknowledgements

D.S.B. thanks Tom Bania for writing the IDL single-dish analysis
package which was used to analyze most of the data in this paper.
Cintia Quireza made sure the observations continued to run smoothly
during some of the early morning periods.  Bob Rood provided
insightful discussion about solar helium abundances.  Lastly,
D.S.B. thank Antonio and Manuel Peimbert for discussions about
temperature fluctuations.

\newpage


\newpage



\begin{deluxetable}{lrcrl}
\small
\tablecolumns{5}
\tablecaption{Hydrogen Spectral Lines}
\tablehead{
\colhead{} & \colhead{$\nu_{\rm rest}$} & \colhead{$\Delta{v}$}
& \colhead{$\nu_{\rm center}$} & \colhead{} \\
\colhead{Transition} & \colhead{(\mhz)$^{\rm a}$} & \colhead{(\kms)$^{\rm b}$}
& \colhead{(\mhz)$^{\rm c}$} & \colhead{Other Transitions$^{\rm d}$}
}
\startdata
H86$\alpha$ & 10161.3029 &  0.360 & 10161.3029 & H108$\beta$; H162$\eta$; H169$\theta$ \\
H87$\alpha$ &  9816.8669 &  0.373 &  9812.0000 & H137$\delta$; H156$\zeta$; H164$\eta$; H171$\theta$ \\
H88$\alpha$ &  9487.8238 &  0.386 &  9505.0000 & H126$\gamma$; H173$\theta$ \\
H89$\alpha$ &  9173.3233 &  0.399 &  9183.0000 & H140$\delta$; H175$\theta$; H188$\kappa$ \\
H90$\alpha$ &  8872.5708 &  0.412 &  8877.0000 & H113$\beta$; H129$\gamma$; H177$\theta$ \\
H91$\alpha$ &  8584.8232 &  0.426 &  8584.8232 & H154$\epsilon$; H179$\theta$; H186$\iota$ \\
H92$\alpha$ &  8309.3850 &  0.440 &  8300.0000 & H132$\gamma$; H145$\delta$; H181$\theta$; H181$\iota$ \\
H93$\alpha$ &  8045.6050 &  0.455 &  8045.6050 & H167$\zeta$; H183$\theta$; H190$\iota$ \\

\enddata
\tablenotetext{a}{Hn$\alpha$ transition rest frequency.}
\tablenotetext{b}{Velocity resolution.}
\tablenotetext{c}{Frequency at the center of the spectrometer 50\mhz\ bandwidth.}
\tablenotetext{d}{Hydrogen transitions within the band for $\Delta{\rm n} \leq\ 10$.}

\label{tab:rrl}
\end{deluxetable}


\begin{deluxetable}{lrccc}
\small
\tablecolumns{5}
\tablecaption{Radio Recombination Line Parameters\tablenotemark{a}}
\tablehead{
\colhead{} & \colhead{$T_{\rm L}$} & \colhead{$\sigma (T_{\rm L})$} &
\colhead{$\Delta v$} & \colhead{$\sigma (\Delta v)$} \\
\colhead{Transition} & \colhead{(mK)} & \colhead{(mK)} &
\colhead{(\kms)} & \colhead{(\kms)}
}
\startdata

H   & 219.12 & 0.14 & 27.83 & 0.02 \\
He  &  23.87 & 0.16 & 21.61 & 0.17 \\
C   &   1.73 & 0.42 &  3.35 & 0.97 \\

\enddata
\tablenotetext{a}{The spectrum consists of the average of the 
$87\alpha - 93\alpha$ transitions, excluding $86\alpha$.  It has an
$rms$ noise of 0.28\mk\ with a total integration time of 555.3\hr.}

\label{tab:results}
\end{deluxetable}

\end{document}